\documentclass[reprint,aps,prd,twocolumn,notitlepage,showpacs,nofootinbib,preprintnumbers,superscriptaddress]{revtex4-1}
\usepackage{amsmath}
\usepackage{amsfonts}
\usepackage{amssymb}
\usepackage{latexsym}
\usepackage{enumerate}
\usepackage{color,xcolor}
\usepackage{graphicx}
\usepackage{bm}
\usepackage{epsfig}
\usepackage[english]{babel}
\usepackage{hyperref}
\usepackage{times}
\usepackage{comment}
\usepackage{epstopdf}

\def\d{\mathrm{d}}
\def\arctanh{\mathrm{arctanh}}
\def\Veff{V_\mathrm{eff}}

\def\risco{r_\mathrm{I}}
\def\phisco{\varphi_\mathrm{I}}

\begin{document}

\title{Innermost stable circular orbit of Kerr-Bertotti-Robinson black holes and inspirals from it: Exact solutions}
\author{Tower Wang}
\email[Electronic address: ]{twang@phy.ecnu.edu.cn}
\affiliation{School of Physics and Electronic Science, East China Normal University, Shanghai 200241, China\\ \vspace{0.2cm}}
\date{\today\\ \vspace{1cm}}
\begin{abstract}
For an uncharged test particle in the Kerr-Bertotti-Robinson spacetime, solutions of two major types of orbits are presented, both in exact forms. First, for both prograde and retrograde motions, the radii of innermost stable circular orbits are expressed fully in terms of the outer and inner horizon radii just in the same form as Kerr black holes, despite the fact that Kerr-Bertotti-Robinson black holes have three parameters. Second, closed analytic solutions are given to the problem of a test particle inspiraling toward the Kerr-Bertotti-Robinson black hole from innermost stable circular orbits at the infinitely distant past. These exact solutions can serve as a springboard for more general solutions and astrophysical applications in the future.
\end{abstract}


\maketitle




It is generally believed that most black holes in the Universe can be described by the Kerr metric \cite{Kerr:1963ud}, a vacuum solution to Einstein equations, discovered almost half a century after Einstein established these equations \cite{Einstein:1915ca}. However, in astrophysical environments, black holes are usually shielded or threaded by a magnetic flux, which is missing from the Kerr metric. In a recent work \cite{Podolsky:2025tle}, Podolsk\'{y} and Ovcharenko presented a novel axisymmetric stationary solution to Einstein-Maxwell equations and dubbed it Kerr-Bertotti-Robinson (KBR) black holes. The solution can be interpreted as a spinning black hole immersed in an external magnetic or electric field that is asymptotically uniform and oriented along the spin axis. It reduces to the Kerr metric when the external field is switched off, and to the Bertotti-Robinson universe \cite{Bertotti:1959pf,Robinson:1959ev} when the black hole mass is set to zero. If further confirmed, this solution will be an important new model for studying black holes immersed in magnetic fields, superior to its counterpart \cite{Wald:1974np,Ernst:1976bsr,DiPinto:2025yaa} in Bonnor-Melvin universe \cite{Bonnor:1954tis,Melvin:1963qx}. Related to this topic, we refer to \cite{Al-Badawi:2004agm,Al-Badawi:2008ucc,Astorino:2022aam,Barrientos:2024pkt,Barrientos:2025rjn,Hu:2025mmp} for a partial list.

In this Letter, we intend to get some insight into KBR black holes by studying timelike geodesics in their equatorial plane. In fact, our principle interest is in orbits of test particles inspiraling from the innermost stable circular orbit (ISCO). Focusing on such orbits, Mummery and Balbus have derived an exact solution of the geodesic equations in the Kerr spacetime \cite{Mummery:2022ana}, laying the foundation for studying accretion flows within the ISCO \cite{Mummery:2023tgh,Mummery:2024rtw,Mummery:2024mrq}. Although it can be reobtained as a special limit of general solutions or integrals reported in Refs. \cite{Mummery:2023hlo,Liu:2023tcy,Cieslik:2023qdc,Ko:2023igf}, the solution for inspirals from the ISCO is dramatically neat because there is a triple root in the effective potential of the radial velocity equation. Along the same line, we will work out the analytic solution of inspiral orbits from the ISCO of KBR black holes. Remarkably, the solution has a closed form. On the half way, we will present the algebraic equation for the ISCO radius of KBR black holes and its solution similar to Kerr black holes. Throughout this Letter, we use geometric units $G=c=1$, where $G$ and $c$ denote the gravitational constant and the light speed, respectively.

According to \cite{Podolsky:2025tle,Ovcharenko:2025cpm}, the gravitational field of the KBR black hole is dictated by the line element
\begin{eqnarray}\label{metric-KBR}
\nonumber\d s^2&=&\frac{1}{\Omega^2}\left\{-\frac{Q}{\rho^2}\left(\d t-a\sin^2\theta\d\varphi\right)^2+\frac{\rho^2}{Q}\d r^2+\frac{\rho^2}{P}\d\theta^2\right.\\
&&\left.+\frac{P}{\rho^2}\sin^2\theta\left[a\d t-\left(r^2+a^2\right)\d\varphi\right]^2\right\},
\end{eqnarray}
where the metric functions are
\begin{eqnarray}
\nonumber\rho^2&=&r^2+a^2\cos^2\theta,\quad Q=(1+B^2r^2)\Delta,\\
\nonumber P&=&1+B^2\left(m^2\,\frac{I_2}{I_1^2}-a^2\right)\cos^2\theta,\\
\nonumber\Omega^2&=&1+B^2r^2-B^2\Delta\cos^2\theta,\\
\Delta&=&\left(1-B^2m^2\frac{I_2}{I_1^2}\right)r^2-2m\frac{I_2}{I_1}\,r+a^2
\end{eqnarray}
with $I_1=1-B^2a^2/2$ and $I_2=1-B^2a^2$. Here $m$ and $a$ are the mass and spin parameters of the black hole, respectively, while $B$ is the asymptotic value of the magnetic (or electric) field. The function $\Delta$ has two zeros \cite{Podolsky:2025tle},
\begin{equation}\label{horizon}
r_\pm=\frac{mI_2\pm\sqrt{m^2I_2-a^2I_1^2}}{I_1^2-B^2m^2I_2} \,I_1,
\end{equation}
which localize the outer and inner black hole horizons. The exact form of electromagnetic field is also available in \cite{Podolsky:2025tle}, but it is irrelevant to our research here. In this Letter, since we are seeking for exact equatorial orbital solutions with closed forms, we do not consider any test particle with an electric or magnetic charge \cite{Shaymatov:2021qvt}, which would deviate from the equatorial plane or induce a solution without a closed form.

In the equatorial plane, or equivalently the timelike hypersurface $\theta=\pi/2$, the nonvanishing components of $g_{\mu\nu}$ and $g^{\mu\nu}$ required for our calculation are as follows:
\begin{eqnarray}
\nonumber&&g_{tt}=\frac{a^2-\bar{\Omega}^2\Delta}{\bar{\Omega}^2r^2},\quad g^{tt}=\frac{a^2\bar{\Omega}^2\Delta-\left(r^2+a^2\right)^2}{r^2\Delta},\\
\nonumber&&g_{rr}=\frac{r^2}{\bar{\Omega}^4\Delta},\quad g_{t\varphi}=g_{\varphi t}=\frac{a\left(\bar{\Omega}^2\Delta-r^2-a^2\right)}{\bar{\Omega}^2r^2},\\
\nonumber&&g^{rr}=\frac{\bar{\Omega}^4\Delta}{r^2},\quad g^{t\varphi}=g^{\varphi t}=\frac{a\left(\bar{\Omega}^2\Delta-r^2-a^2\right)}{r^2\Delta},\\
&&g_{\varphi\varphi}=\frac{\left(r^2+a^2\right)^2-a^2\bar{\Omega}^2\Delta}{\bar{\Omega}^2r^2},\quad g^{\varphi\varphi}=\frac{\bar{\Omega}^2\Delta-a^2}{r^2\Delta},
\end{eqnarray}
in which $\bar{\Omega}^2=1+B^2r^2$. For a particle of rest mass $\mu$ moving in this hypersurface, the Hamiltonian takes the form
\begin{equation}\label{Ham}
\mathcal{H}=\frac{1}{2}\left(g^{tt}p_t^2+g^{rr}p_r^2+2g^{t\varphi}p_tp_\varphi+g^{\varphi\varphi}p_\varphi^2\right),
\end{equation}
which is subject to the constraint $\mathcal{H}=-\mu^2/2$. This constraint is equivalent to the normalization condition
\begin{equation}\label{unorm}
g^{tt}u_t^2+g^{rr}u_r^2+2g^{t\varphi}u_tu_\varphi+g^{\varphi\varphi}u_\varphi^2=-1,
\end{equation}
because the contravariant velocity $u^{\alpha}\equiv\d x^{\alpha}/\d\tau$ is related to the contravariant momentum by $u^{\alpha}=p^{\alpha}/\mu$. Here $\tau$ denotes the proper time, and we have chosen the affine parameter as $\tau/\mu$. The Hamiltonian \eqref{Ham} does not depend explicitly on coordinates $t$ and $\varphi$, whose conjugate momenta are thereby conserved. As a result, we find two constants of motion: the energy per unit mass $\varepsilon=-u_t=-p_t/\mu$, and the angular momentum per unit mass $\ell=u_{\varphi}=p_{\varphi}/\mu$. Inserting them into \eqref{unorm} gives
\begin{equation}
\left(u^r\right)^2+g^{rr}\left(1+g^{tt}\varepsilon^2-2g^{t\varphi}\varepsilon\ell+g^{\varphi\varphi}\ell^2\right)=0.
\end{equation}
This equation is of the form $\left(u^r\right)^2+\Veff=0$, where
\begin{eqnarray}
\nonumber\Veff&=&g^{rr}\left(1+g^{tt}\varepsilon^2-2g^{t\varphi}\varepsilon\ell+g^{\varphi\varphi}\ell^2\right)=-\frac{\bar{\Omega}^4}{r^4}R(r),\\
R(r)&\equiv&\left(\varepsilon r^2+\varepsilon a^2-\ell a\right)^2-\Delta\left[r^2+\bar{\Omega}^2\left(\ell-\varepsilon a\right)^2\right].\label{Rr}
\end{eqnarray}
In terms of the conserved quantities, the remaining components of the contravariant velocity can be written as
\begin{eqnarray}
\nonumber u^t&=&-g^{tt}\varepsilon+g^{t\varphi}\ell=\frac{1}{r^2}T(r),\\
T(r)&\equiv&a\bar{\Omega}^2\left(\ell-\varepsilon a\right)+\frac{r^2+a^2}{\Delta}\left(\varepsilon r^2+\varepsilon a^2-\ell a\right),\\
\nonumber u^{\varphi}&=&-g^{\varphi t}\varepsilon+g^{\varphi\varphi}\ell=\frac{1}{r^2}\Phi(r),\\
\Phi(r)&\equiv&\bar{\Omega}^2\left(\ell-\varepsilon a\right)+\frac{a}{\Delta}\left(\varepsilon r^2+\varepsilon a^2-\ell a\right).
\end{eqnarray}

For particles moving along an ISCO of radius $r=\risco$, the radial potential satisfies $\Veff(\risco)=\Veff'(\risco)=\Veff''(\risco)=0$. They are equivalent to  $R(\risco)=R'(\risco)=R''(\risco)=0$ and can be regarded as constraints on $\risco$, $\varepsilon$ and $\ell$. It is thus clear that $r=\risco$ is a triple root of $R(r)=0$. In addition, it can be checked directly that the constant term is exactly zero in the fourth-order polynomial $R(r)$, resembling the effective radial potential of Kerr black holes \cite{Mummery:2022ana}. To reproduce the expected roots, we write down $R(r)=-R_0\,r(r-\risco)^3$ with $R_0$ being an undetermined coefficient. Comparing it with \eqref{Rr} order by order and making some manipulations, we find
\begin{eqnarray}\label{varepsilon}
\nonumber &&R_0=\frac{2m}{\risco\left(3-B^2\risco^2\right)}\frac{I_2}{I_1},\quad\left(\ell-\varepsilon a\right)^2=\frac{\risco^2}{3-B^2\risco^2},\\
&&\varepsilon^2=\frac{1}{3-B^2\risco^2}\left(3-3B^2m^2\frac{I_2}{I_1^2}-\frac{2m}{\risco}\frac{I_2}{I_1}\right),\\
\nonumber&&\ell^2=\frac{1}{3-B^2\risco^2}\left[\left(\risco^2-3a^2\right)B^2m^2\frac{I_2}{I_1^2}+\left(3\risco^2-a^2\right)\frac{2m}{\risco}\frac{I_2}{I_1}\right].
\end{eqnarray}
It is useful to rearrange them as
\begin{eqnarray}
\nonumber2\varepsilon a\ell&=&\frac{1}{3-B^2\risco^2}\left[3a^2-\risco^2+\left(\risco^2-6a^2\right)B^2m^2\frac{I_2}{I_1^2}\right.\\
&&\left.+\left(3\risco^2-2a^2\right)\frac{2m}{\risco}\frac{I_2}{I_1}\right].
\end{eqnarray}
Then the ISCO radius $\risco$ can be evaluated according to $(2\varepsilon a\ell)^2=4\varepsilon^2a^2\ell^2$, which turns out to be a quartic equation
\begin{eqnarray}
\nonumber&&\left(I_1^2-B^2m^2I_2\right)^2\left(\frac{\risco}{I_1}\right)^4-12mI_2\left(I_1^2-B^2m^2I_2\right)\left(\frac{\risco}{I_1}\right)^3\\
\nonumber&&+6\left[m^2I_2\left(6I_2+B^2a^2\right)-a^2I_1^2\right]\left(\frac{\risco}{I_1}\right)^2-28ma^2I_2\left(\frac{\risco}{I_1}\right)\\
&&+9 a^4=0.
\end{eqnarray}
In order to get $\risco$ from this equation, let us rewrite it as
\begin{eqnarray}
\nonumber&&\risco^4-6(r_++r_-)\risco^3+\left[9(r_++r_-)^2-6r_+r_-\right]\risco^2\\
&&-14r_+r_-(r_++r_-)\risco+9r_+^2r_-^2=0.
\end{eqnarray}
To our surprise, the coefficients in this equation are determined exclusively by the radii of outer and inner horizons, in spite of the fact that the KBR black hole possesses three parameters.\footnote{But we became less surprised when we noticed the finding in \cite{Podolsky:2025tle} that the ISCO radius of Schwarzschild-Bertotti-Robinson black holes is three times the horizon radius. That corresponds to the $\lambda=0$ limit of \eqref{ISCO} in the $a=0$ limit of \eqref{metric-KBR}.} The equation should also hold for Kerr black holes exactly. Therefore, as long as the ISCO radii of Kerr black holes mentioned in \cite{Wang:2025vsx} is reliable, we can infer that the ISCO radii of KBR black holes ought to be
\begin{eqnarray}\label{ISCO}
\nonumber\risco&=&\frac{r_++r_-}{2}\left[3+Z_2\mp\sqrt{(3-Z_1)(Z_1+2Z_2+3)}\right],\\
\nonumber Z_1&=&1+\left(1-\lambda^2\right)^{1/3}\left[\left(1+\lambda\right)^{1/3}+\left(1-\lambda\right)^{1/3}\right],\\
Z_2&=&\sqrt{3\lambda^2+Z_1^2},
\end{eqnarray}
where we have introduced a dimensionless parameter $\lambda=2\sqrt{r_+r_-}/(r_++r_-)$ for convenience. The upper (or lower) sign corresponds to prograde (or retrograde) orbits defined by $2\varepsilon a\ell>0$ (or $2\varepsilon a\ell<0$).

From aforementioned radial and angular equations of motion, it is straightforward to write down the differential orbit equation $(\d r/\d\varphi)^2=R(r)/\Phi(r)^2$. We are interested in particles inspiraling from the ISCO, whose orbit is determined by the solution
\begin{equation}\label{phir}
\varphi-\phisco=-\int_{\risco}^{r}\frac{\Phi(r)}{\bar{\Omega}^2\sqrt{R(r)}}\d r
\end{equation}
in the parameter space $\risco\geq r\geq r_+\geq r_-\geq 0$, with $\risco$ being the radius of ISCO, and $\phisco$ being the initial value of azimuthal angle. We will return to the value of $\phisco$ soon, but for the moment let us pay attention to the right hand side of \eqref{phir}, which takes the concrete form
\begin{widetext}
\begin{eqnarray}\label{insp}
\nonumber\varphi-\phisco&=&-\int_{\risco}^{r}\frac{\ell-\varepsilon a}{\sqrt{R_0\,r(\risco-r)^3}}\d r-\int_{\risco}^{r}\frac{a\left(\varepsilon r^2+\varepsilon a^2-\ell a\right)}{\left(1+B^2r^2\right)\left(1-B^2m^2\frac{I_2}{I_1^2}\right)(r-r_-)(r-r_+)\sqrt{R_0\,r(\risco-r)^3}}\d r\\
\nonumber&=&C_0\left.\sqrt{\frac{r}{\risco-r}}\,\right|_{\risco}^r+C_+\arctanh\frac{\sqrt{r_+(\risco-r)}}{\sqrt{r(\risco-r_+)}}+C_-\arctanh\frac{\sqrt{r_-(\risco-r)}}{\sqrt{r(\risco-r_-)}}\\
&&+\frac{C_1(D_0+D_2)}{\sqrt{2B_+}}\arctan\frac{\sqrt{2B_+r(\risco-r)}}{\risco-B_+r}+\frac{C_1(D_0-D_2)}{\sqrt{2B_-}}\arctanh\frac{\sqrt{2B_-r(\risco-r)}}{\risco+B_-r},\\
\nonumber C_0&=&-\frac{2\left(\ell-\varepsilon a\right)}{\risco\sqrt{R_0}}
-\frac{a}{\risco\sqrt{R_0}\left(1-B^2m^2\frac{I_2}{I_1^2}\right)}
\frac{2\left(\varepsilon\risco^2+\varepsilon a^2-\ell a\right)}{\left(1+B^2\risco^2\right)(\risco-r_-)(\risco-r_+)},\\
\nonumber C_\pm&=&\frac{a}{\sqrt{R_0}\left(1-B^2m^2\frac{I_2}{I_1^2}\right)}
\frac{2\left(\varepsilon r_\pm^2+\varepsilon a^2-\ell a\right)}{\left(1+B^2r_\pm^2\right)(r_\pm-r_\mp)\sqrt{r_\pm(\risco-r_\pm)^3}},\\
\nonumber C_1&=&\frac{a}{\sqrt{R_0}\left(1-B^2m^2\frac{I_2}{I_1^2}\right)}\frac{B^2\left[B^2\left(\varepsilon a^2-\ell a\right)-\varepsilon\right]}{\left(1+B^2\risco^2\right)\left(1+B^2r_-^2\right)\left(1+B^2r_+^2\right)},\quad B_\pm=\sqrt{1+B^2\risco^2}\pm1,\\
D_2&=&\frac{2\risco\left(1-B^2r_+r_-\right)+\left(1-B^2\risco^2\right)(r_++r_-)}{\sqrt{1+B^2\risco^2}},\quad D_0=\risco\left(1-B^2r_+r_-\right)+(r_++r_-).
\end{eqnarray}
\end{widetext}
The main tricks in deriving \eqref{insp} are the substitution $r=\risco u^2/(1+u^2)$ with $u>0$, and the indefinite integrals
\begin{eqnarray}
\nonumber\int\frac{\d u}{b^2u^4+2u^2+1}&=&\frac{1}{2\sqrt{2(b+1)}}\arctan\frac{\sqrt{2(b+1)}\,u}{1-bu^2}\\
\nonumber&+&\frac{1}{2\sqrt{2(b-1)}}\arctanh\frac{\sqrt{2(b-1)}\,u}{1+bu^2},\\
\nonumber\int\frac{u^2\d u}{b^2u^4+2u^2+1}&=&\frac{1}{2b\sqrt{2(b+1)}}\arctan\frac{\sqrt{2(b+1)}\,u}{1-bu^2}\\
\nonumber&-&\frac{1}{2b\sqrt{2(b-1)}}\arctanh\frac{\sqrt{2(b-1)}\,u}{1+bu^2}\\
\end{eqnarray}
with a constant $b>1$. Although cumbersome, \eqref{insp} is an elementary function of $r$. Therefore, the solution of inspiral orbits starting from the ISCO of KBR black holes has a closed form. Here the specific energy $\varepsilon$ and angular momentum $\ell$ take the values in \eqref{varepsilon}, while $r_\pm$ and $\risco$ are dictated by \eqref{horizon} and \eqref{ISCO}. The initial value of azimuthal angle is arbitrary in principle, but it would be convenient to rid off constant terms on both sides of \eqref{insp} by setting $\phisco$ to the lower limit of the first term on the right hand side.

In the limit $r_+=r_-$ for the extremal black hole, one can see from \eqref{ISCO} directly that the ISCO radius $\risco=r_+$ for prograde orbits, and $\risco=9r_+$ for retrograde orbits. In this limit, most terms in \eqref{insp} are regular, but the coefficients $C_\pm$ are divergent, thus the second and third terms in \eqref{insp} require extra attention. Combing the two terms together and taking the limit $r_+=r_-$ carefully, we find the divergence cancels. Eventually, in the extremal limit, the sum of the two terms with coefficients $C_\pm$ are found to be
\begin{eqnarray}
\nonumber&&\frac{a}{\sqrt{R_0}\left(1-B^2m^2\frac{I_2}{I_1^2}\right)}
\frac{2\left(\varepsilon r_+^2+\varepsilon a^2-\ell a\right)}{\left(1+B^2r_+^2\right)\sqrt{r_+(\risco-r_+)^3}}\\
\nonumber&&\times\left\{\left[\frac{2\varepsilon r_+}{\varepsilon r_+^2+\varepsilon a^2-\ell a}-\frac{2B^2r_+}{1+B^2r_+^2}-\frac{\risco-4r_+}{2r_+(\risco-r_+)}\right]\right.\\
\nonumber&&\left.\times\arctanh\frac{\sqrt{r_+(\risco-r)}}{\sqrt{r(\risco-r_+)}}+\frac{\sqrt{r(\risco-r)}}{2(r-r_+)\sqrt{r_+(\risco-r_+)}}\right\}.\\
\end{eqnarray}

In summary, we have shown that the ISCO radius of KBR black holes has the same closed form as the one of Kerr black holes in terms of the outer and inner horizon radii, and that the orbit equation for inspirals starting from ISCO of KBR black holes has an exact solution. They are given by \eqref{ISCO} and \eqref{insp} as the main results of this Letter. Given the importance of these orbits in black hole physics and the importance of spinning black holes in the magnetic field \cite{Wang:2025vsx,Jai-akson:2017ldo,Zeng:2025olq,Zeng:2025tji}, we believe that our results will be valuable for studying more general solutions and astrophysical applications. It should be intriguing to investigate how generally the first result holds for other black holes and its implications to a holographic interpretation of ISCOs in quantum gravity \cite{Festuccia:2008zx,Berenstein:2020vlp}.

\acknowledgments{This work is sponsored by the Natural Science Foundation of Shanghai (Grant No. 24ZR1419300).}


\end{document}